# Asymptotic Close to Optimal Resource Allocation in Centralized Multi-band Wireless Networks

Mohammadreza Darabi, Amin Roustaei
Tipco, *Tehran, Iran*
Sales4@Tipcon.ir , Sales1@Tipco.ir

*Abstract*— This paper concerns sub-channel allocation in multi-user wireless networks with a view to increasing the network throughput. It is assumed there are some sub-channels to be equally divided among active links, such that the total sum rate increases, where it is assumed each link is subject to a maximum transmit power constraint. This problem is found to be a non-convex optimization problem and is hard to deal with for large number of sub channels and/or users. However, relying on some approximation methods, it is demonstrated that the proposed sub-optimal problem has roots in combinatorial optimization, termed as Assignment problem which can be tackled through the so called Hungarian method. Simulation results demonstrate that the proposed method outperforms existing works addressed in the literature.

I. INTRODUCTION

Resource allocation is one of the main challenges in wireless networks. This is due to the limited amount of bandwidth, stringent power constraint and more demands for emerging communication services. Orthogonal Frequency-Division Multiple Access, i.e., OFDMA, is an effective method to divide the available bandwidth into some orthogonal subcarriers in a multi-user multi-path dispersive channel. In such systems the allocation problem is abstracted to effectively assign subcarriers to current users/links as well as finding the best power allocation strategy across subcarriers such that a cost function is optimized. In this regard, some efforts are carried out to address proper resource allocation in such networks which can be abstracted in two main categories: fixed resource allocation [1] versus dynamic resource allocation [2], [3].

In fixed resource allocation, available resources are distributed among users based on a predefined algorithm regardless of the channel status of each user. In dynamic resource allocation, however, the Channel State Information (CSI) is used to effectively allocate existing resources. As a result, dynamic resource allocation yields better performance at the expense of imposing more complexity to the network as the channel information should be available at a centralized node to effectively assign sub-channels to communication links.

Dynamic resource allocation can be divided into two main subclasses: (i) Margin Adaptive (MA) where the objective is to achieve the minimum transmit power when each user has a specific rate constraint [4], (ii) and Rate Adaptive (RA) where the objective is to maximize the sum-rate capacity when there is a stringent transmit power constraint [2]. Also, there are some other works emphasizing on minimizing the BER or maintaining a quality of service [5].

This paper studies RA problem in a multi-carrier wireless network where the problem is to equally distribute subcarriers among users such that the network throughput increases. This problem is known to be NP-hard in general, thus it does not have a trivial solution. However, relying on some approximation methods, it is demonstrated that the new problem can be translated to a known combinatorial problem, called Assignment problem which can be tackled through the use of the so-called Hungarian method.

The rest of this paper is organized as follows. Section II presents the system model followed by the problem formulation. Section III presents the proposed resource allocation. Section IV illustrates numerical results and finally Section V concludes the paper with findings.

II. System Model and Problem Statement

In this paper, we consider a multi-band communication network encompassing $K$ communication links and $N$ sub-channels, where equal number of sub-channels, i.e., $\lfloor \frac{N}{K} \rfloor$, should be allocated to each link. Also, it is assumed that the channel information associated with each link is available. The main objective persuaded in the current work is to allocate equal number of sub-channels to each active link such that network throughput is maximized, assuming each link is subject to a peak transmit power constraint. This problem can be cast as the following optimization problem,

$$\max_{p_{k,n}, \rho_{k,n}} \frac{B}{N} \sum_{k=1}^{K} \sum_{n=1}^{N} \rho_{k,n} \log_2 \left( 1 + \frac{p_{k,n} |h_{k,n}|^2}{N_0 \frac{B}{N}} \right)$$

$$s.t. \sum_{n=1}^{N} p_{k,n} \leq P_K \text{ for all } k$$

$$p_{k,n} \geq 0 \text{ for all } n, k$$

$$\rho_{k,n} = \{0,1\} \text{ for all } n, k$$

$$\sum_{k=1}^{K} \rho_{k,n} = 1 \text{ for all } n$$

(1)

where $N_0$ is the noise power spectral density; $B$ and $P_K$ represent the total bandwidth and the available power for each link, respectively. Also, $h_{k,n}$ and $p_{k,n}$ are, respectively, the channel gain associate with the $k$th link in the $n$th sub-channel and the allocated power and $\rho_{k,n}$ is a zero-one indicator representing whether the $n$th sub-channel is assigned to the $k$th link or not, ensuring each sub-channel is assigned to just one communication link.

One can readily verify that the optimization problem in (1) involves some continuous variables $p_{k,n}$ as well as binary variables $\rho_{k,n}$ which is not convex in general, thus it does not yield a trivial solution. This motivated us to pursue addressing the aforementioned problem through relying on two approximations which are, respectively, tight for high and low SNR regimes. Then, the simplified problem is tackled in two steps: (i) attempting to assign best sub-channels to each link and (ii) finding the best power allocation strategy across sub-channels.

III. Resource Allocation

This section aims at finding the best sub-channels as well as optimum power allocation strategy across sub-channels at two marginal cases of low and high SNR regimes. To this end, referring to (1), we note that for the $i$th link, the achievable rate becomes,

$$R_i = \frac{B}{N} \sum_{n \in \pi(i)} \log_2(1 + p_{i,n} H_{i,n})$$
$$s.t. \sum_{n \in \pi(i)} p_{i,n} \leq P_i \quad (2)$$

where $\pi(i)$ is the set of sub-channels assigned to the $i$th link which is of size $\left\lfloor \frac{N}{K} \right\rfloor$. Also, $H_{i,n}$ is set to $H_{i,n} = \frac{|h_{i,n}|^2}{N_0 \frac{B}{N}}$.

A. Low SNR regime

In this case, noting $p_{i,n} H_{i,n} \ll 1$ and $\log_2(1 + x) \approx \frac{1}{\ln 2} x$, the achievable rate of the $i$th user can be approximated as,

$$R_i = \frac{B}{N \ln 2} \sum_{n \in \pi(i)} p_{i,n} H_{i,n}$$
$$s.t. \sum_{n \in \pi(i)} p_{i,n} \leq P_i \quad (3)$$

Knowing the set $\pi(i)$ and assuming $n_i = \arg\max_{n \in \pi(i)} H_{i,n}$, one can readily observe that $R_i$ is maximized when total available power is assigned to the best sub-channel in terms of having maximum gain, i.e., $R_i^{max} \approx \frac{B}{N \ln 2} P_i H_{i,n_i}$. As a result, the maximum sum-rate throughput becomes

$$R^{max} = \sum_{i=1}^{K} R_i^{max} = \frac{B}{N \ln 2} \sum_{i=1}^{K} P_i H_{i,n_i} \quad (4)$$

In other words, for computing (4), one just need to know the best sub-channel in each set $\pi(i)$ for $i = 1, ..., K$. This problem can be tackled through the so called assignment problem. In a typical assignment problem there are $K$ machines and $K$ jobs, where the cost of doing the $j$th job on the $i$th machine is $c_{ij}$. The problem is to effectively assign theses jobs to available machines such that the total cost decreases (or the benefit increases). The Hungarian method is shown to solve the assignment problem in a polynomial time [6]. As a result, the cost matrix $C$ associated with the assignment problem is constructed such that the element of the $i$th row and the $j$th column, i.e., $c_{ij}$, is set to $c_{ij} = P_i H_{i,j}$, thus Hungarian method finds the best sub-channel assignment which maximizes $\sum_{i=1}^{K} \sum_{j=1}^{N} \rho_{ij} c_{ij}$, where $\rho_{ij}$ is a zero/one indicator ensuring each sub-channel is assigned to at most one link, i.e., $\sum_{i=1}^{K} \rho_{ij} = 1$.

*B. High SNR regime*

In this case, we note that $p_{i,n} H_{i,n} \gg 1$ is much greater than one for the selected sub-channels, implying $\log(1 + p_{i,n} H_{i,n}) \approx \log(p_{i,n} H_{i,n})$. As a result, (3) can be well approximated as,

$$R_i \approx \frac{B}{N} \sum_{n \in \pi(i)} \log(p_{i,n} H_{i,n}) = \frac{B}{N} \log\left(\prod_{n \in \pi(i)} p_{i,n}\right) + \frac{B}{N} \log\left(\prod_{n \in \pi(i)} H_{i,n}\right) \quad (5)$$
$$s.t. \sum_{n \in \pi(i)} p_{i,n} \leq P_i$$

One can readily verify that for the selected sub-channels $H_{i,n}$, the best power allocation strategy, i.e., maximizing $\prod_{n \in \pi(i)} p_{i,n}$ under the constraint $\sum_{n \in \pi(i)} p_{i,n} \leq P_i$, gives equal power for each sub-channel, i.e., $p_{i,n} = \frac{P_i}{\left\lfloor \frac{N}{K} \right\rfloor}$, where it is assumed $\pi(i)$ is of size $\left\lfloor \frac{N}{K} \right\rfloor$. However, the set of best sub-channels, i.e., $\pi(i)$ for $i = 1, ..., K$, have yet to be addressed. To this end, referring to (5), we note that $\log(\prod_{n \in \pi(i)} H_{i,n}) = \sum_{n \in \pi(i)} \log(H_{i,n})$. As a result, setting the $i$th row and the $j$th column of cost matrix $C$ to $c_{ij} = \log(H_{i,j})$, it turns out that the corresponding assignment problem maximizes $\sum_{i=1}^{K} \sum_{j=1}^{N} \rho_{ij} c_{ij}$, where again $\rho_{ij}$ is a zero/one indicator ensuring each sub-channel is assigned to at most one link, i.e., $\sum_{i=1}^{K} \rho_{ij} = 1$.

IV. SIMULATION RESULTS

For the sake of comparison with the optimal sub-channel assignment, since the brute-force algorithm should be employed to derive the optimal solution, we concentrate on the special case of having $K = 2$ links and $N = 4$ sub-channels, where it is assumed the channel gains are Rayleigh distributed with zeros mean and unit variance which may experience shadowing with probability $P_f = 0.02$. Moreover, we assume that $\frac{B}{N} = 1$ and $N_0 = 1$. For the optimal sub-channel assignment, the water-filling power allocation strategy is being used to allocate the available power across sub-channels. The results are also compared to the so-called max-select method addressed in [7] in which best sub-channels in terms of having maximum channel strengths are recursively selected in a greedy fashion.

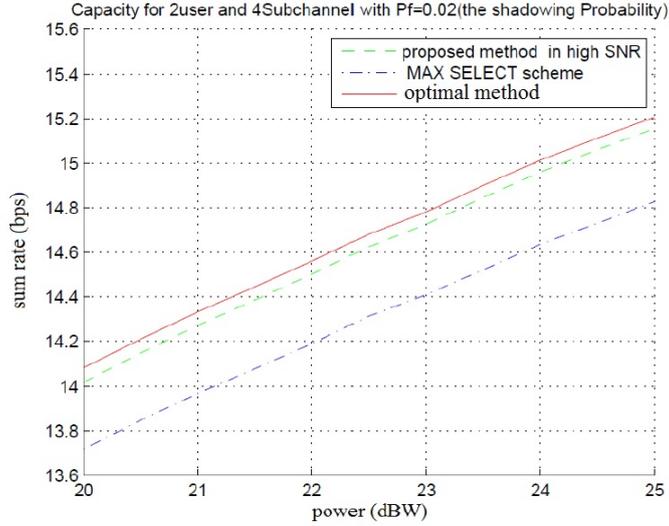

Fig. 1. $N = 4$ and $K = 2$, sum-rate capacity versus power budget for high SNR

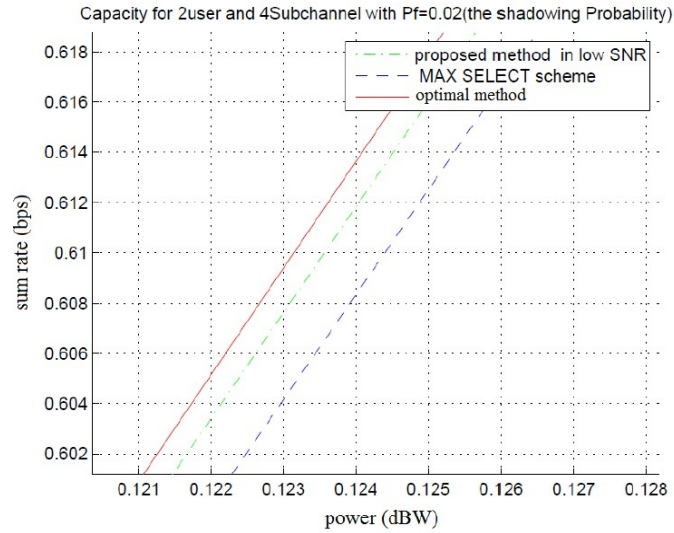

Fig. 2. $N = 4$ and $K = 2$, sum-rate capacity versus power budget for low SNR

Fig. 1 and Fig. 2 illustrate the sum-rate capacity versus power budget, indicating at low and high SNR regimes, the proposed methods coincides the optimal method.

From Complexity viewpoint, the computational complexity of the proposed method, optimal method and max-select method are $\mathcal{O}(N^4)$, $\mathcal{O}(KN^K)$ and $\mathcal{O}(N)$, respectively. It is seen that the proposed method has less complexity than optimal method for high values of $N$ and $K$.

V. CONCLUSION

This paper concerns Rate Adaptive problem for multi-band centralized wireless networks. Since the optimal solution is found to be computationally infeasible, two close to optimal strategies for low and high SNR regimes are proposed, showing there is a close agreement between them with the optimal solution in these regions. Moreover, the proposed methods are low complex, hence, they have some practical implications.